\documentclass[aps,prx,reprint,superscriptaddress,showpacs]{revtex4-1}
\usepackage{color}
\usepackage{graphicx}
\usepackage[american]{babel}
\usepackage[T1]{fontenc}
\usepackage{amsmath}
\usepackage{amssymb}
\usepackage{amstext}
\usepackage{amsthm}
\usepackage{latexsym}
\usepackage{verbatim}
\usepackage{natbib}
\usepackage{array}
\usepackage{color} 

\usepackage{ulem}   

\newcommand{\be}{\begin{equation}}
\newcommand{\ee}{\end{equation}}
\newcommand{\bea}{\begin{eqnarray}}
\newcommand{\eea}{\end{eqnarray}}

\newcount\bozza \bozza=0
\ifnum\bozza=1
\newdimen\shift \shift=-2truecm
\def\lb#1{%
{\label{#1}\rlap{\kern\shift{$\scriptstyle#1$}}}}
\else\def\lb#1{\label{#1}} \fi

\begin{document}

\title{Influence of the surface symmetry breaking on the magnetism, collapsing and three dimensional dispersion of Co pnictides ACo$_2$As$_2$ (A=Ba, Sr, Ca)}

\author{Joseph Mansart}
\affiliation{Laboratoire de Physique des Solides, Universit\'{e} Paris-Sud, UMR 8502, B\^at. 510, 91405 Orsay, France}

\author{Patrick Le F\`evre}

\author{Fran\c{c}ois Bertran}
\affiliation{Synchrotron SOLEIL, L'Orme des Merisiers, Saint-Aubin-BP 48, 91192 Gif sur Yvette, France}

\author{Anne Forget}

\author{Doroth\'{e}e Colson}
\affiliation{Service de Physique de l'Etat Condens\'{e}, Orme des Merisiers, CEA Saclay, CNRS-URA 2464, 91191 Gif sur Yvette Cedex, France}
\author{V\'eronique Brouet $^{\dagger,} $}
\affiliation{ Laboratoire de Physique des Solides, Universit\'{e} Paris-Sud, UMR 8502, B\^at. 510, 91405 Orsay, France}

\begin{abstract}
We use angle-resolved photoemission (ARPES) to study the three dimensional (3D) electronic structure of Co pnictides ACo$_2$As$_2$ with A=Ba, Sr, Ca or a mixture of Sr and Ca. These compounds are isostructural to Fe based superconductors, but have one more electron in the Co $3d$ orbitals. Going from Ba to Ca, they become more and more 3D, eventually forming a \lq\lq{}collapsed\rq\rq{} tetragonal phase, where the distance between CoAs layers is markedly reduced. We observe with ARPES the periodicity of the electronic structure as a function of $k_z$ (i.e. perpendicularly to CoAs layers) and find that it matches in each case that expected from the distance between the planes in the bulk. However, the electronic structure is better fitted by a calculation corresponding to a slab with 2 CoAs layers than to the bulk structure. We attribute this to subtle modifications of the 2D electronic structure induced by the truncation of the 3D dispersion at the surface in the ARPES measurement. We further study how this affects the electronic properties. We show that, despite this distortion, the electronic structure of CaCo$_2$As$_2$ is essentially that expected for a collapsed phase. Electronic correlations produce a renormalization of the electronic structure  by a factor 1.4, which is not affected by the transition to the collapsed state. On the other hand, a small shift of the Fermi level reduces the density of states in the $e_g$ bands and suppresses the magnetic transition expected in CaCo$_2$As$_2$. Our study evidences that observing the 3D bulk periodicity is not sufficient to ensure bulk sensitivity. It further gives direct information on the role of 3D interactions, mostly governed by Co-As hybridization, among $e_g$ and $t_{2g}$ orbitals. It is also useful to better understand the electronic structure of Fe superconductors and the range of validity of ARPES measurements. 
\end{abstract}

\date{\today}

\maketitle

\section{Introduction}

Layered systems play a central role in the study of correlated systems, since the reduced dimensionality typically enhances the strength of electronic  correlations. Cuprates, iron-based superconductors, transition-metal chalcogenides are all examples of layered systems. While the electronic structure of one layer often works as a good approximation for the electronic structure of the bulk material, interlayer couplings are ultimately also crucial. This is evidenced by default by the specific properties discovered more and more in monolayers, from graphene to MoS$_2$. The most direct probe of the electronic structure of one system is arguably angle-resolved photoemission (ARPES). However, it is extremely surface sensitive, the typical escape depth of electrons at the photon energies enabling high resolution (10-100eV) being 5\AA  \cite{Hufner,TejedaMalterre}.~One can therefore question whether the measured electronic structure corresponds to that of the bulk, or the surface, or a monolayer and how these different cases connect to each other. 

We tackle this problem by investigating a family of Co pnictides, ACo$_2$As$_2$, where the distance between the CoAs layers can be tuned from 5.2\AA~to 6.3\AA~by intercalating different ions A (Ba, Sr, Ca) between the layers. This family is isostructural to the \lq\lq{}122\rq\rq{} family of Fe superconductors, as sketched in Fig. 1a. For A=Ca, the structure is \lq\lq{}collapsed\rq\rq{}, as found in CaFe$_2$As$_2$ under pressure \cite{KreyssigPRB08}, meaning that, although the structural symmetry remains the same, the distance between CoAs layers is markedly reduced, enabling the formation of As dimers between the planes (see arrows in Fig. 1a). We establish that ARPES measures very well the bulk distance between the CoAs layers, but also evidence traces of an underlying bilayer structure. We investigate the meaning of this finding by detailed comparison between measurements and calculations for bulk and 2 layers slab structures. 

\begin{figure*}[tb]
\centering
\includegraphics[width=0.9\textwidth]{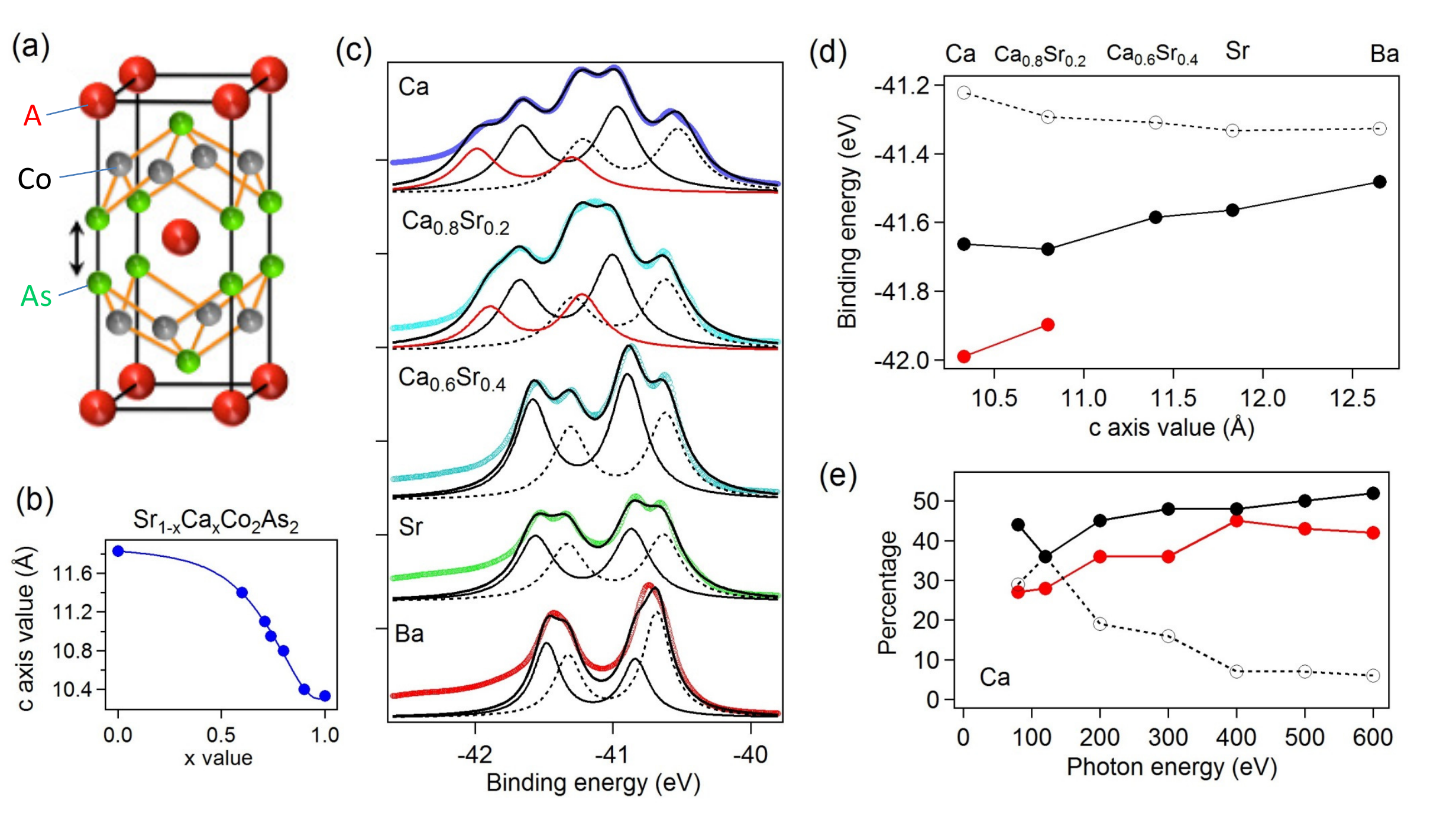}
\caption{(a) Sketch of the structure of ACo$_2$As$_2$. (b) c axis value as a function of the nominal fraction of Ca and Sr. (c) Photoemission spectra showing the As $3d$ core levels measured at 100~eV photon energy in the indicated compounds. Each As gives rise to a doublet due to spin-orbit splitting and up to three As doublets are observed as indicated by dotted, solid and red lines. (d) Peak positions of the lowest binding energy for each As $3d$ doublet as a function of c. (e) Percentage of the three doublets as a function of photon energy for CaCo.}
\label{As_CL}
\end{figure*}

These Co pnictides are isostructural with the \lq\lq{}122\rq\rq{} family of Fe superconductors \cite{PaglioneReview}, which exhibit a very rich multi-orbital electronic structure, enabling original types of electronic correlations \cite{YinNatureMat11,Demedici14}. When Fe is fully substituted with Co, there is one more electron per site, i.e. 7 electrons in the $3d$ shell. The general expectation is that Co compounds will be less correlated than their Fe counterparts \cite{XuDingPRX13}, because they are further away from half filling, where correlations would be the strongest. However, an interesting feature is that the extra electron reaches $e_g$ levels that are empty for Fe pnictides. They form a very flat band near the Fermi level, yielding a high density of states at the Fermi level $n(E_F)$, favorable to Stoner-like ferromagnetism \cite{SefatPRB09,PandeyPRB13,AnandPRB14_Ca}. Co pnictides then offer the opportunity to study the physics of this structure in a more itinerant limit than Fe pnictides, where magnetism is also present but probably of a very different nature. 

In BaCo$_2$As$_2$ \cite{SefatPRB09} and SrCo$_2$As$_2$ \cite{PandeyPRB13}, there is an enhanced magnetic susceptibility, suggesting proximity of a quantum instability, but no magnetic transition down to low temperatures. In CaCo$_2$As$_2$, on the other hand, there is a transition to an antiferromagnetic (AF) state below 72K \cite{ChengPRB12,YingPRB12}, where the CoAs layers order ferromagnetically \cite{QuirinalePRB13}, in good agreement with the ferromagnetic tendency expected from the high $n(E_F)$. The case of Ba is the one that has been studied by ARPES in more details \cite{XuDingPRX13,DhakaPRB13}, although some data are also available for Sr \cite{PandeyPRB13} and Ca \cite{AnandPRB14_Ca}. In this paper, we use ARPES to give a complete picture of the electronic structure of these compounds. We evidence an increased 3D character from Ba to Ca, in good agreement with bulk calculations (section III). There are, however, some deviations that are better explained by a 2 layers slab structure calculation (section IV). We finally give special attention to the transition in CaCo$_2$As$_2$ to the collapsed structure (section V) and magnetic phase (section VI). We conclude that the structure measured by ARPES for Ca is collapsed, but not magnetic. This is attributed to an upward shift of the $e_g$ levels at the surface, which proves by default their role in the magnetic transition. We finally discuss the implications of these findings for ARPES studies of Fe pnictides in general (section VII).

\begin{figure*}[tbh]
\centering
\includegraphics[width=0.9\textwidth]{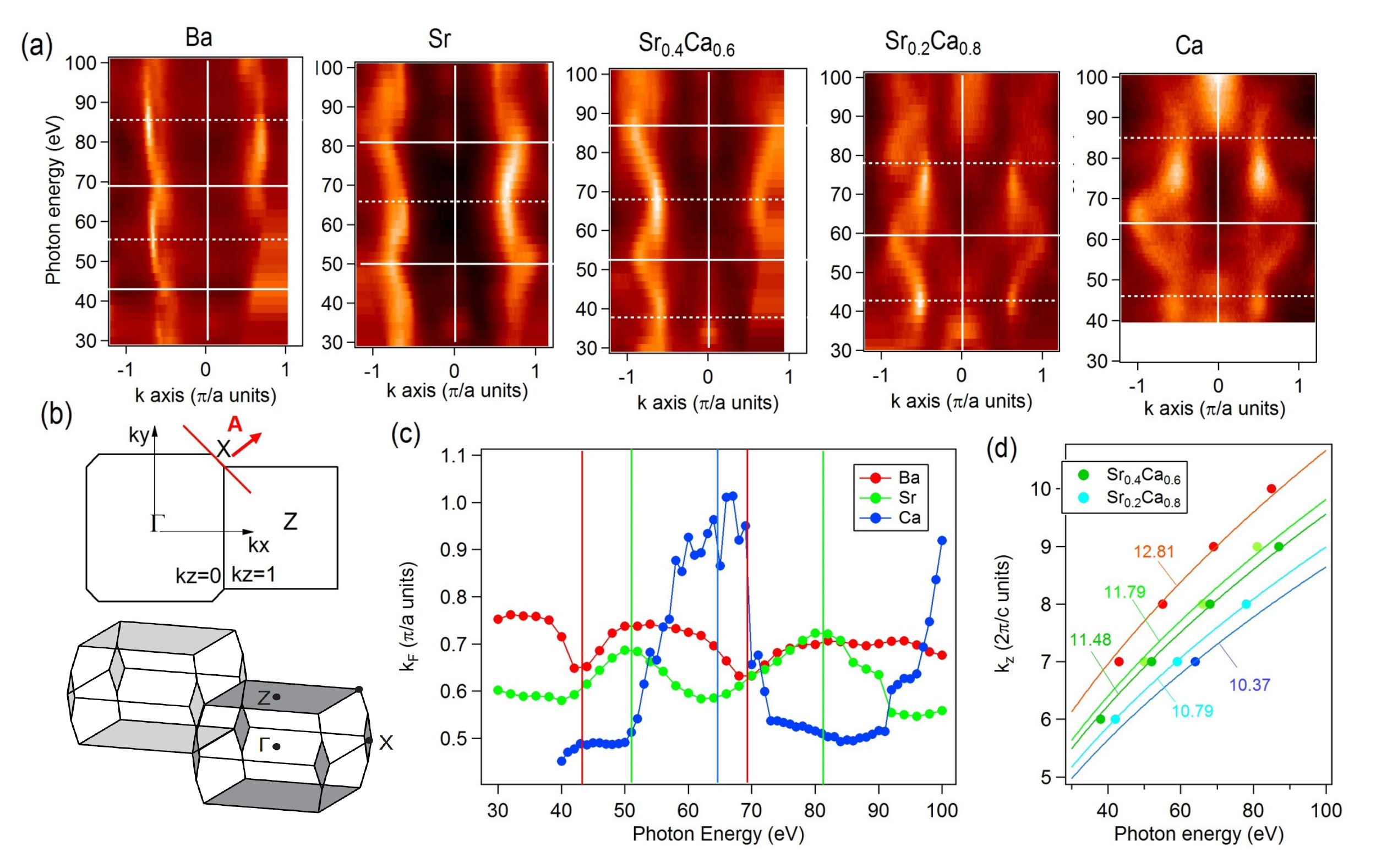}
\caption{(a) Spectral weight integrated in a 10 meV window around $E_F$ in the indicated compounds, as a function of photon energy. The k axis is indicated in (b). X is indicated by a white vertical line. White horizontal lines indicate the extrema corresponding to even (solid) or odd (dotted) values of $k_z$. (b) Sketch of two adjacent BZ at $k_z$=0 and $k_z$=1 (top) and 3D view (bottom). The red line indicates the direction of the cut measured in (a) using s-polarization. (c) $k_F$, extracted from the previous maps, as a function of photon energy for BaCo, SrCo and CaCo. Extrema corresponding to odd $k_z$ are indicated by vertical lines. (d) Extrema of $k_z$, in units of 4$\pi$/c, determined experimentally as a function of photon energy for the different compounds. Solid lines are fits to Eq. (2) assuming V$_0$=13.5eV and giving the values of c indicated on the graph.}
\label{Periodicity}
\end{figure*}

\section{Experimental and calculation details}

We have synthesized single crystals of ACo$_2$As$_2$ (abbreviated as ACo in the following), for A=Ba, Sr, Ca, with methods similar to those used in Fe pnictides previously \cite{RullierAlbenquePRL09}. Starting reagents of high purity Ba, Sr and/or Ca and CoAs were mixed in the molar ratio 1:4. For ARPES measurements, the crystals were cleaved in ultra-high vacuum at low temperatures. The cleavage plane is between two CoAs layers (see Fig. 1a) and the surface neutrality is preserved by removing half the A ions of the top layer \cite{MasseePRB09}. To better study the transition between \lq\lq{}normal\rq\rq{} and \lq\lq{}collapsed\rq\rq{} phases \cite{KreyssigPRB08}, we synthesized compounds with intermediate values of c using a combination of Ca and Sr. The evolution of c is given in Fig. \ref{As_CL}(b). The magnetic properties of the low Sr doping were studied in \cite{YingPRB12} and a similar structural evolution was reported for (Ca,Sr)Co$_2$P$_2$ \cite{JiaPRB09}. In this paper, we report on Ca$_{0.8}$Sr$_{0.2}$ (c=10.82\AA) and Ca$_{0.6}$Sr$_{0.4}$ (c=11.41\AA). In our CaCo sample, the magnetic transition, measured by SQUID, takes place at 72K, as reported by \cite{ChengPRB12}, but different from ref. \cite{AnandPRB14_Ca}, where T$_N$=52K was obtained and the stoichiometry was estimated to CaCo$_{1.86}$As$_2$. 

The $3d$ core-levels of As, presented in Fig. \ref{As_CL}(c), allow to evidence the transition to the collapsed phase. They can generally be fitted by a pair of spin-orbit split doublets, assigned to bulk (solid black lines) and surface (dotted black lines) contributions. In the collapsed phase, a new doublet appears at high binding energies, showed in red, probably as a result of a highest electronic density around As in this structure. We observed a similar shift to high energy in collapsed CaFe$_2$As$_2$ (not shown), as also reported in \cite{TsubotaJPhysSocJapan13}. This evidences that the last two compounds have collapsed structures. All these peak positions are reported in Fig. \ref{As_CL}(d). Experiments at high photon energy, up to 600eV, in Fig. \ref{As_CL}(e), demonstrate that the first two doublets are bulk in nature, while the outer one is a surface contribution. The reason why three doublets appear in the collapsed phase is still to be determined.

Our ARPES experiments were carried out at the CASSIOPEE beamline of the SOLEIL synchrotron, with a Scienta R4000 analyzer, an angular resolution of 0.3$^{\circ}$ and an overall energy resolution from 5meV around 30eV to 15meV around 100eV. For most of the data shown in the paper, the polarization A was Linear Vertical (LV), lying in the sample surface plane. This selects even orbitals in the plane containing A and the surface normal (this will be called p-polarization in the following) and odd orbitals for the perpendicular plane (s-polarization). All measurements were done at 20K, except when noted otherwise. 

We performed band structure calculations using the WIEN2k software \cite{Wien2k}. We used structures determined experimentally for BaCo (c=12.65\AA) \cite{AnandPRB14_Ba}, SrCo (c=11.8\AA) \cite{PandeyPRB13} and CaCo (c=10.33\AA) \cite{JohnstonAdvPhys10}. Interestingly, in all these cases, the position $z$ of As above the Co plane is $\sim$ 1.27\AA,~very close from the value we optimized theoretically. Note that early determination of $z$ \cite{JohnstonAdvPhys10} gave the opposite impression with $z$ decreasing from 1.41\AA~in BaCo to 1.27\AA~in CaCo. This is at variance with the case of Fe pnictides, where $z$ is usually much larger in experiment than the calculated one. This is believed to be the sign of the presence of local magnetic moments in Fe pnictides, as the optimized value of $z$ in magnetic calculations correspond much better to the experimental one \cite{MazinPRB08}. This can be taken a contrario as an indirect proof that such moments have indeed disappeared in all Co pnictides. 

\section{Three dimensional electronic structure}
We start our analysis by investigating the periodicity of the $k_z$ dispersion, perpendicularly to the cleaved surface. This can be done with ARPES by scanning the photon energy h$\nu$. The conservation of energy during the photoemission process \cite{Hufner,TejedaMalterre} dictates that the kinetic energy $E_{kin}$ of the detected electron is that of the photon minus the binding energy $E_B$ of the photoexcited electron and the work function W paid to cross the surface.

%
\begin{align}
E_{kin}&=h\nu-W-E_B\\
           &=\hbar^2(k_{//}^2+k_{z}^2)/2m-V_0
\end{align}
\label{Eq1}

\begin{figure}[tbh]
\centering
\includegraphics[width=0.5\textwidth]{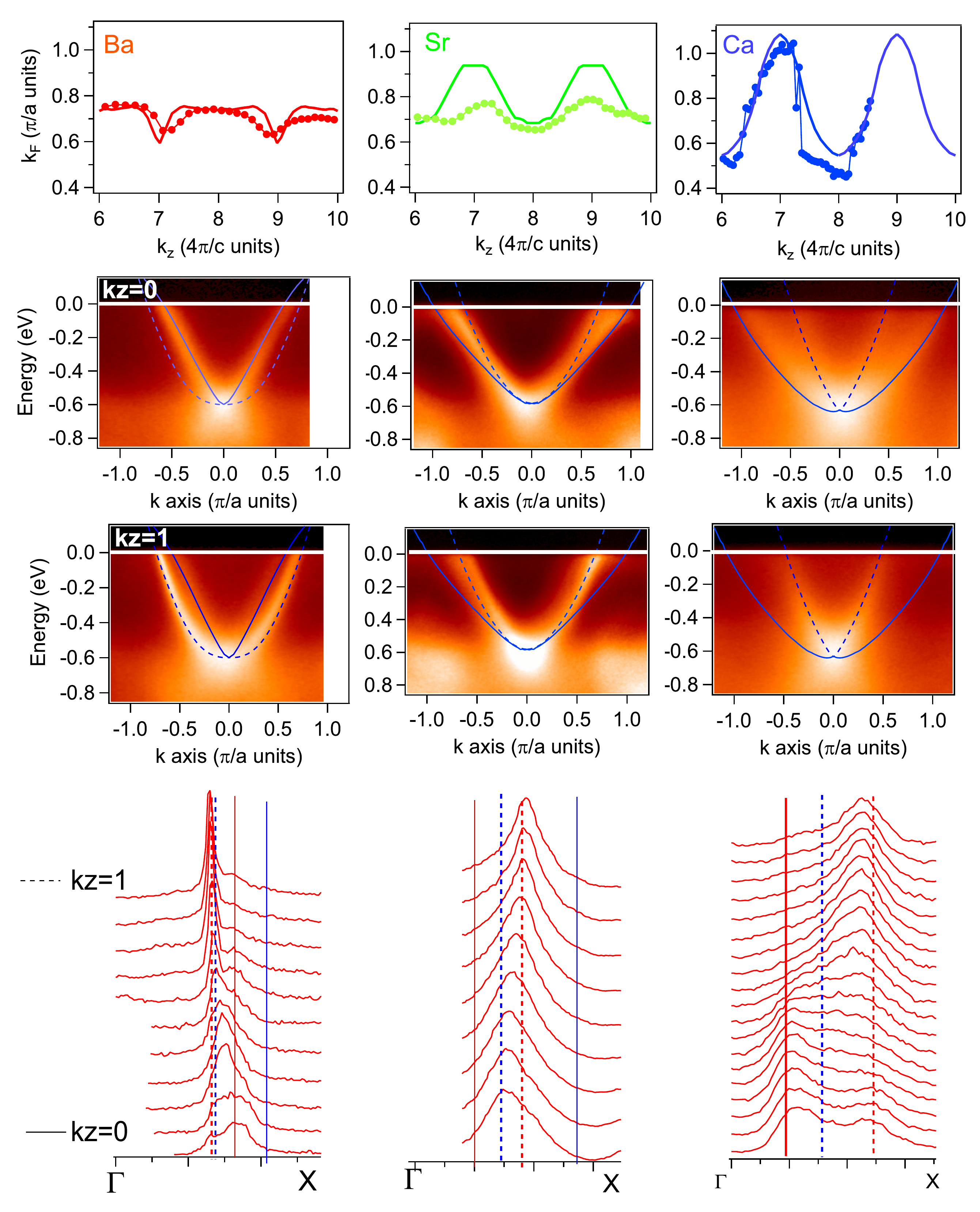}
\caption{The different columns correspond to data measred for BaCo, SrCo and CaCo, respectively. 1st row : symbols are experimental values of $k_F$ as a function of $k_z$ for the $d_{xz}$/$d_{yz}$ electron band. They were extracted from the maps of Fig. 2 by lorentzian fits and/or determination of local maxima. The contour expected in bulk calculation is shown as solid line. 2nd row : Energy-momentum images of dispersion along $\Gamma$X at $k_z$=0 (68eV for BaCo, 50eV for SrCo and 63eV for CaCo). $k=0$ corresponds to X. Lines indicate bulk calculation at $k_z$=0 (solid line) and $k_z$=1 (dotted line) renormalized by 1.4. For CaCo, data are also shifted up by 70~meV. 3rd row : Same for $k_z$=1 (86eV for BaCo, 68eV for SrCo and 85eV for CaCo). 4th row : MDC stacks along $\Gamma$X from $k_z$=0 (lower spectrum) to $k_z$=1 (higher spectrum). Vertical lines indicate the calculated positions of $k_F$ in bulk calculation at $k_z$=0 (solid) and $k_z$=1 (dotted) for $d_{xz}$/$d_{yz}$ (red) and $d_{xy}$ (blue).}
\label{VsKz}
\end{figure}
\begin{figure*}[tbh]
\centering
\includegraphics[width=0.95\textwidth]{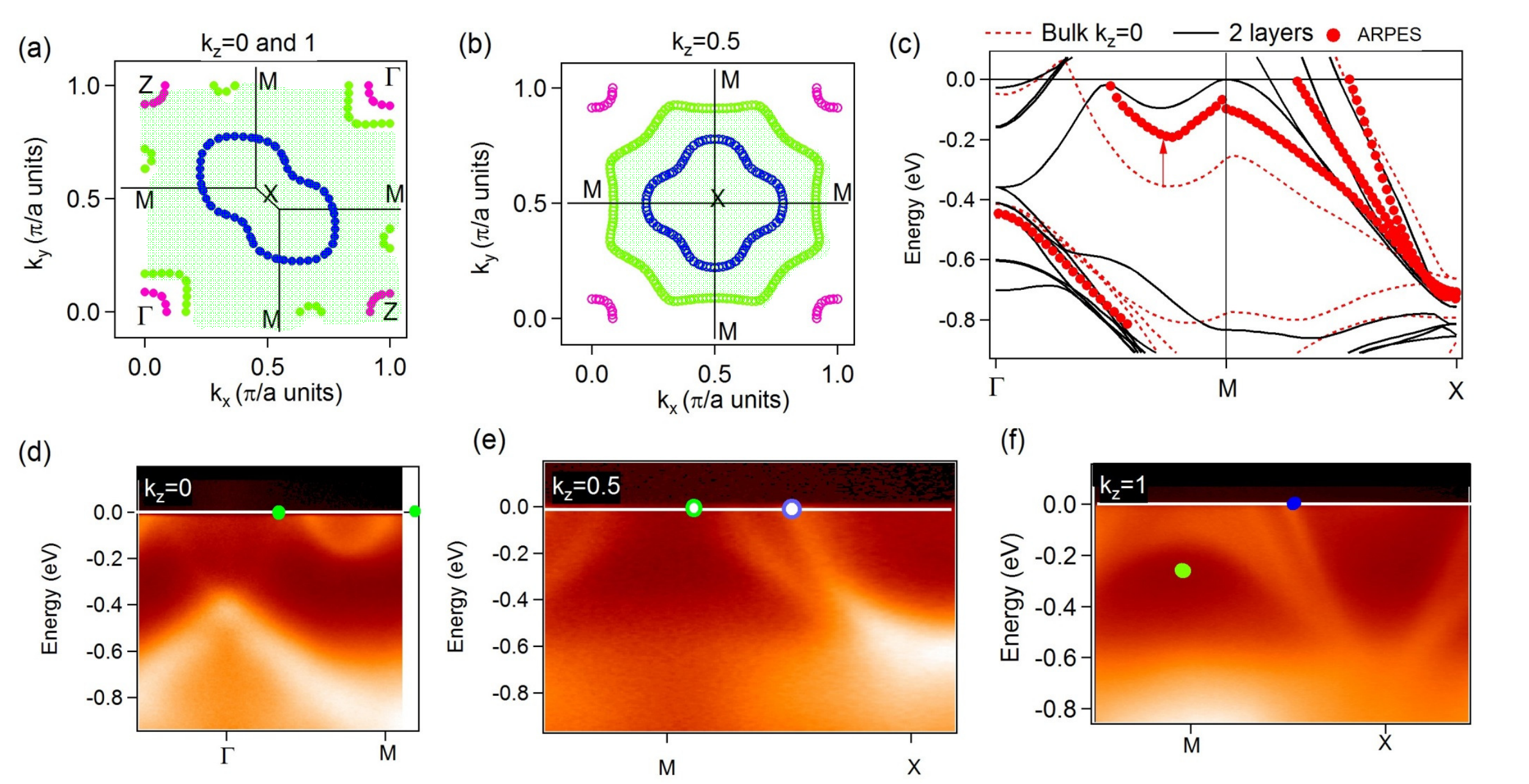}
\caption{(a-b) Calculated FS for the bulk SrCo at $k_z$=0 and 1 (a) and $k_z$=0.5 (b). Colors indicate the predominant orbital character : $d_{x^2-y^2}$ (pink), $d_{xy}$ (blue) or $d_{xz}$/$d_{yz}$ (green). The filled part of the green sheet is shaded in green. (c) Red circles : experimental dispersion along $\Gamma$MX, taken from the images shown in (d-f). They are compared with calculation for the bulk at $k_z$=0 (red dotted line) and for the 2 layers slab model (black lines), both renormalized by 1.4. (d) Energy-momentum image measured at 30eV ($k_z$=0) along $\Gamma$M. Green points indicate FS crossings expected from (a). (e) Energy-momentum image measured at 70eV ($k_z$=0.5) along MX. Open green and blue points indicate FS crossings expected from (b). (f) Energy-momentum image measured at 44eV ($k_z$=1) along MX. Blue point indicates FS crossing expected from (a) and green point the position of the band forming the green sheet at M in (a). }
\label{SrCo}
\end{figure*}

\noindent This outgoing electron is usually assumed to be in a free electron state. It is defined by a parabola with a bottom at an inner potential -$V_0$, as written in Eq. 2, where $k_{//}$ and $k_{z}$ are components of the momentum of the electron before crossing the surface. The wave vector parallel to the surface $k_{//}$ is conserved, but not the component perpendicular to the surface $k_{z}$. However, one sees from Eq. (1) and (2) that $k_{z}$ will increase with $h\nu$ and one expects periodic variations in units of 2$\pi$/c for a fixed $k_{//}$. This is how $k_z$ will be determined. For these compounds, the expected 3D periodicity along $k_z$ is 4$\pi$/c, as there are 2 planes per unit cell.

In Fig. \ref{Periodicity}(a), we show the variation with the photon energy of the spectral weight integrated at the Fermi level, along the diagonal of the Brillouin Zone (BZ). This direction is shown in Fig. \ref{Periodicity}(b) as a red line and the polarization A, perpendicular to it, selects odd orbitals, in this case mainly the $d_{xz}/d_{yz}$ electron band \cite{BrouetLiFeAs}. Note that the stacking of the BZ is such that when X lies in the plane defined by $k_z$, this direction measures a dispersion corresponding to 1-$k_z$ (see Fig. 2b). In all cases, there is a clear periodic variation with photon energy. The successive extrema are marked with white solid/dotted lines for even/odd $k_z$. In Fig. \ref{Periodicity}(c), we plot $k_F$ as a function of photon energy extracted from these maps to evidence how these extrema are shifting in the different compounds. 

These different periodicities correspond to the different values of c. Usually, $k_z$ is adjusted from this periodicity using, in Eq. (2), the bulk value of c and fitting $V_0$ as a parameter. Here, we fix $V_0$=13.5~eV, a usual value for Fe pnictides \cite{BrouetPRB09,MalaebJPSJ09}, and fit the values of c. We obtain a good description of the data, as shown in Fig. \ref{Periodicity}(d), and values of c strikingly close to the bulk values, as indicated on the graph. This shows unambiguously that ARPES is sensitive to the bulk distance between the planes and that there is no strong variation of this distance near the surface. 

Next, we replot in Fig. \ref{VsKz} these contours as a function of $k_z$. They match quite remarkably the contours calculated for the bulk structures. For example, we observe an inversion of minima and maxima at $k_z$=1 between BaCo and SrCo, as calculated, and a large increase of the $k_z$ dispersion between SrCo and CaCo, corresponding to the transition to the collapsed phase. Nevertheless, in SrCo, the $k_z$ dispersion appears smaller experimentally than theoretically. 

In rows 2 and 3 of Fig. \ref{VsKz}, we show the energy-momentum images at $k_z$=0 and 1, together with the bulk calculation for $d_{xz}$/$d_{yz}$, renormalized by 1.4 and shifted up by 70~meV for CaCo. Although a $d_{xy}$ electron band is also expected in this direction, it is not intense in these experimental conditions and will be studied later. The calculated dispersion describes generally quite well the experiment. In particular, the bottom of the band, which is fixed in $k_z$, is well fitted in each case with this renormalization. The renormalization value is in good agreement with the one previously found for BaCo \cite{XuDingPRX13}. It is smaller than the one used for BaFe$_2$As$_2$ lightly doped with Co (we estimated $\sim$2 for electron pockets in ref. \cite{BrouetPRL13}), but not dramatically despite the large change of doping. Moreover, it does not seem to evolve much with the increased dimensionality and/or the collapsed transition in SrCo and CaCo.

For BaCo, the bulk calculation reproduces very well the characteristic difference of band shapes from triangular at $k_z$=0 to rounded at $k_z$=1. For SrCo and CaCo, there is a good agreement at $k_z$=1, but it is not as good at $k_z$=0. For SrCo, the band does not open as much as expected, as already noticed, and, f or CaCo, it seems that we observe simultaneously dispersions corresponding to $k_z$=0 and $k_z$=1. A closer inspection of the MDC lineshapes, displayed in the 4th row of Fig. \ref{VsKz}, shows that there is a strong broadening as a function of $k_z$ in all compounds, although it is more obvious for CaCo. The lineshape are rather good lorentzians near $k_z$=1, but distorts for smaller $k_z$. The intensity tends to spread between the values expected at $k_z$=0 (solid red line) and 1 (dotted red line), often with local maxima at these two extreme values. We also indicate with blue lines the positions expected for $d_{xy}$, which evidences these lineshapes cannot be explained by a residual contribution from $d_{xy}$. The most dramatic case is CaCo at $k_z$=0, where there is almost a \lq\lq{}doublet\rq\rq{} structure corresponding to these two values. We observed a similar broadening with $k_z$ in LiFeAs \cite{BrouetLiFeAs}, so that we believe these lineshapes are typical from Fe pnictides. It is of course crucial to take this into account to attempt any fitting of the dispersions and avoid mistaking this structure for two different bands. 

A broadening of the MDC lineshape in 3D systems is commonly expected in ARPES, as the incertitude on k$_{z}$ induces an integration over a finite $k_z$ range \cite{StrocovJEleSpec03,DerondeauCondMat16}. The lineshape should depend both on the $k_z$ dispersion and the lifetime of the final state \cite{BansilPRB05}, so that a detailed prediction is difficult. However, the doublet observed here at $k_z$=0 is also reminiscent of a bilayer system. In systems with few layers, one expects to build progressively a 3D dispersion, when adding more and more layers. This was nicely observed for a few layers of graphene \cite{OhtaPRL07}, where N peaks were observed for N layers, with oscillating intensities as a function of photon energy. Such oscillations are expected from the different symmetries of the wave functions (bonding and antibonding for a bilayer system \cite{MoreschiniPRB14}). As N increases, the dispersion of the N peaks converge to the 3D dispersion. This raises questions on the meaning of the 3D dispersion observed here, especially whether thinking in terms of an underlying bilayer structure could be meaningful.

\begin{figure*}[tbh]
\centering
\includegraphics[width=0.9\textwidth]{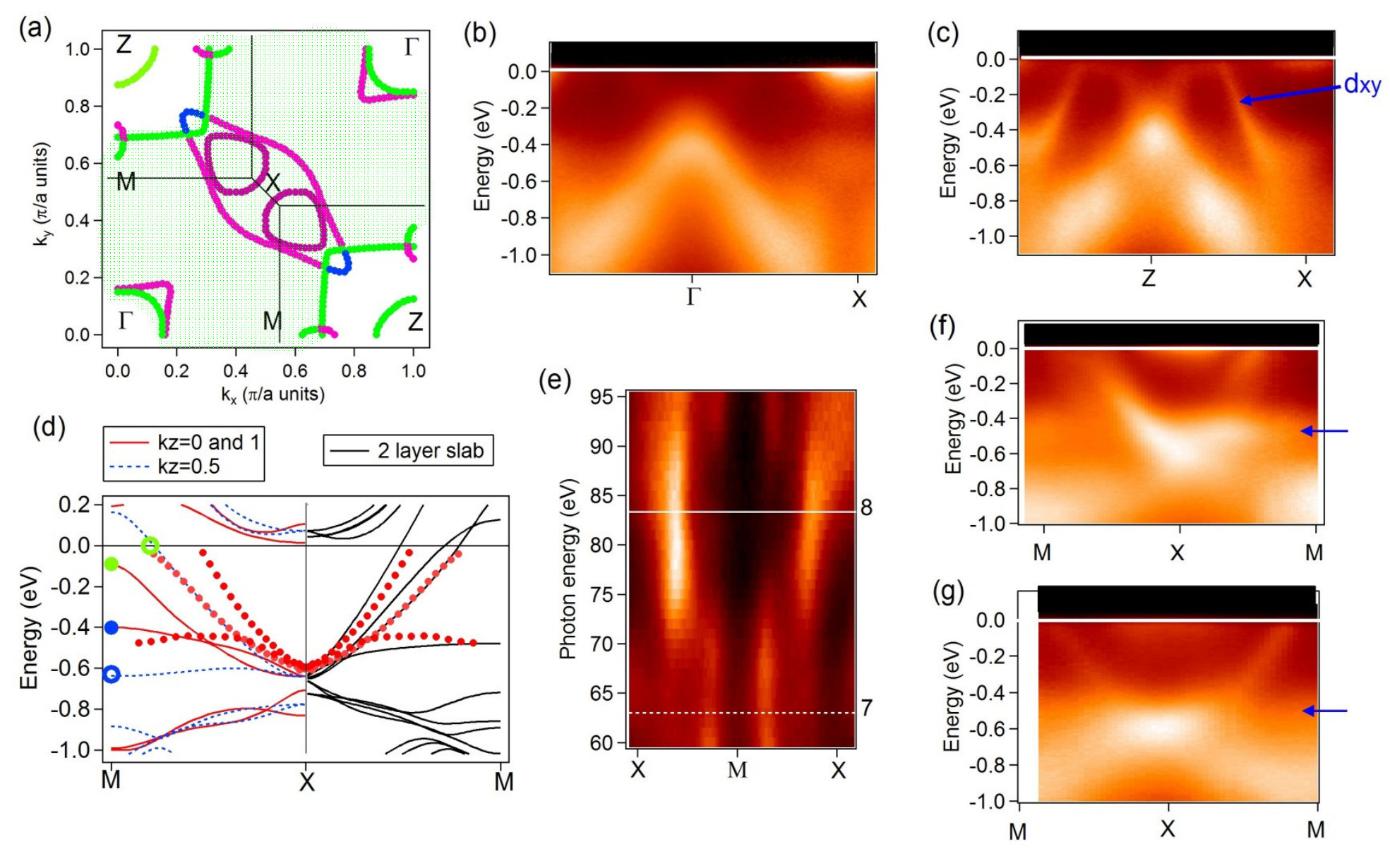}
\caption{(a) Fermi Surface calculated for CaCo at $k_z$=0 and 1. The main orbital character of the different sheets is $e_{g}$ for pink and purple dots, $d_{xz}$/$d_{yz}$ for green dots, $d_{xy}$ for blue dots. The occupied part of the green sheet is shaded in green. (b) Dispersion along the BZ diagonal at 82~eV ($k_z$$\sim$0) with p-polarization. (c) Same at 106~eV ($k_z$$\sim$1). (d) Red circles : experimental dispersion obtain near $k_z$=0 and 1, compared on the left with bulk calculation at $k_z$=0 (red lines) and 0.5 (blue dotted lines) and on the right with the 2 layers slab calculation (black lines). In both cases, the calculation is renormalized by 1.4 and shifted up by 70~meV. (e) Spectral weight integrated at the Fermi level along MX direction as a function of photon energy. (f) Dispersion along MX at 82~eV ($k_z$$\sim$0). (g) Dispersion along MX at 100~eV ($k_z$$\sim$0.5). The blue arrows indicate the position of the flat band indicated by blue circles in (d).}
\label{CaCo}
\end{figure*}

\section{Model with 2 layers slab}

To explore this idea, we have performed calculations for a 2 layers slab. To avoid doubling the unit cell, we consider the structure of Fig. \ref{As_CL}(a) with 7\AA~of vaccum above and below and remove one electron to the alkali-earth atom facing vacuum, to preserve stoichiometry. In reality, stoichiometry is preserved by removing half of the alkali-earth at the surface \cite{MasseePRB09}. We keep the structural positions to the bulk values. We have observed that relaxing these positions modifies only marginally the electronic structure. We believe this is a minimal model to explore how the symmetry breaking at the surface impacts the electronic structure. 

The major impact of the 3D dispersion occurs in fact at the limits of the BZ. As exemplified in Fig. \ref{SrCo}(a-b) in the case of SrCo, the FS consists of small electron pockets of $d_{x^2-y^2}$ character around the $\Gamma$ point (pink color) and two larger sheets of predominantly $d_{xy}$ (blue) and $d_{xz}$/$d_{yz}$ (green) character. The area filled in green indicates the filled parts of the green sheet. At $k_z$=0.5, the two sheets form large electron pockets around X, as they do in Fe pnictides, but, at $k_z$=0 and 1, the outer sheet is large enough to intersect the BZ boundaries. Along $\Gamma$M, it forms a small electron pocket, delimited by green points on Fig. \ref{SrCo}(d). We indeed observe such a pocket experimentally, although it is a bit smaller in diameter. Along MX, at $k_z$=0.5 [Fig. \ref{SrCo}(e)], two electron pockets are expected (blue and green points), as observed, although again with slightly different $k_F$ values. Along MX at $k_z$=0 or 1 [Fig. \ref{SrCo}(f)], only one band should cross the Fermi level (blue point), the other band rising up to -0.2eV at M (green point). We observe a clear band crossing $E_F$ at the blue point, but, for the other band, there is a large distribution of intensities starting well above the green point and extending to the Fermi level. 

For a quantitative comparison, we report in Fig. \ref{SrCo}(c) these experimental dispersions (red circles) together with the bulk calculation at $k_z$=0 (red dotted line) and the 2 layers slab calculation (black lines). All calculations are renormalized by 1.4, as for Fig. \ref{VsKz}. While this describes reasonably well most features, there is a clear deviation near M, where the calculated bulk band should be shifted up by 0.2~eV to fit the experiment (see arrow). Interestingly, the deviation between the bulk and 2 layers slab calculation exhibits a similar behavior, with the largest deviation in this area. We conclude that the deviation observed in SrCo is indeed due to a deviation between the 3D dispersion measured by ARPES and the 3D bulk dispersion. We suggest that comparing bulk and few layers calculation is a relatively easy way to pinpoint these deviations.

\section{Transition to the collapsed phase}

Having evidenced that the electronic structure measured by ARPES may deviate from the bulk structure, it is interesting to wonder whether ARPES will be sensitive to the transition to the collapsed phase that depends crucially on interaction between As of neighboring layers. Indeed, As dimers will be broken at the cleaved surface, so that, despite the fact that we know from Fig. \ref{Periodicity} that the distance between the CoAs layers is that of the collapsed phase in our CaCo sample, some modification could occur. Fig. \ref{CaCo}(a) gives the FS expected from bulk calculation at $k_z$=0 and 1. The main difference compared to SrCo is that the $d_{xy}$ electron sheet is largely suppressed, except for a small portion near $k_z$=1. This was explained in details by Andersen and Boeri \cite{Andersen2011}. As a counter part, a large pocket of $d_{x^2-y^2}$/$d_{z^2}$ character develops around X. 

Our data support the large suppression of the $d_{xy}$ electron band. To observe it, we measure again along the diagonal of the BZ, but with the light polarization along this direction, instead of perpendicularly to it, as in Fig. \ref{Periodicity}(b) \cite{BrouetLiFeAs}. At $k_z$=1, in Fig. \ref{CaCo}(c), we observe $d_{xy}$ clearly, as marked by the arrow. However, it rapidly disappears away from that $k_z$ value. At $k_z$=0, shown in Fig. \ref{CaCo}(b), there is no clear sign of it. 

The band structure along MX gives definitive evidence of the transition to the collapsed phase. On the left part of Fig.\ref{CaCo}(d), we show the bulk band calculation at $k_z$=0 and $k_z$=0.5, for the collapsed CaCo phase, renormalized by 1.4 and shifted up by 70~meV, as in Fig. \ref{VsKz}. As a consequence of the suppression of the blue sheet, two bands (marked by blue circles) remain below $E_F$ at $k_z$=0. One is nearly flat reaching -0.4eV at M. In Fig. \ref{CaCo}(f-g), we observe such a flat band (see arrow), supporting the transition to the collapsed phase, as was also suggested from the change in the As core-levels (Fig. \ref{As_CL}) or the larger $k_z$ dispersion of $d_{xz}$/$d_{yz}$ (Fig. \ref{VsKz}). Thus, the broken As dimers at the surface do not prevent the transition to the collapsed phase. Interestingly, this is also true for the 2 layer slab calculation [right part of Fig. \ref{CaCo}(d)], where one band has strongly shifted down compared to SrCo [see Fig. \ref{SrCo}(c)]. However, there are still 3 bands approaching $E_F$, instead of only 2 in the bulk, suggesting a smaller degree of collapsing. These differences in the degree of collapsing are probably responsible for the shift of 70~meV we have to apply to the bulk calculation to fit our data in CaCo. The $d_{xy}$ electron sheet may, for example, contain some more electrons than in the fully collapsed case, requiring a reduction of the size of the green sheet. 

As for SrCo, there are also deviations between the measured bands and the bulk calculation, where the hybridization with bands having strong 3D dispersion is the largest, i.e. near the M point. In our data, we observe for all $k_z$ one or two bands crossing the Fermi level along MX, as seen on the map as a function of photon energy of Fig. \ref{CaCo}(e), whereas only one around $k_z$=0.5 is expected in bulk calculation. This means that the green sheet does not extend as much in the BZ as expected in the bulk calculation. As in Fig. \ref{Periodicity}, there is a characteristic beating of intensity between two extrema, with coexistence for some photon energies. As $k_z$=0 and $k_z$=1 are equivalent by symmetry along MX, it is clear here that it is not due to a different band structure, but just to a modulation of the intensities. The images corresponding to the smallest and largest $k_F$ are shown in Fig. \ref{CaCo}(f-g). Their dispersions are reported in Fig. \ref{CaCo}(d) and it is clear that the calculated bulk bands (left part) should be shifted up at M by ~0.2eV to fit the data. On the other hand, the 2 layers slab calculation (right part) is in fair agreement with the measurement. Again, we conclude that the truncated 3D dispersion at the surface modifies the 2D dispersion of the bands in the plane, especially near M. 

\section{Magnetic phase}

\begin{figure}[tbh]
\centering
\includegraphics[width=0.5\textwidth]{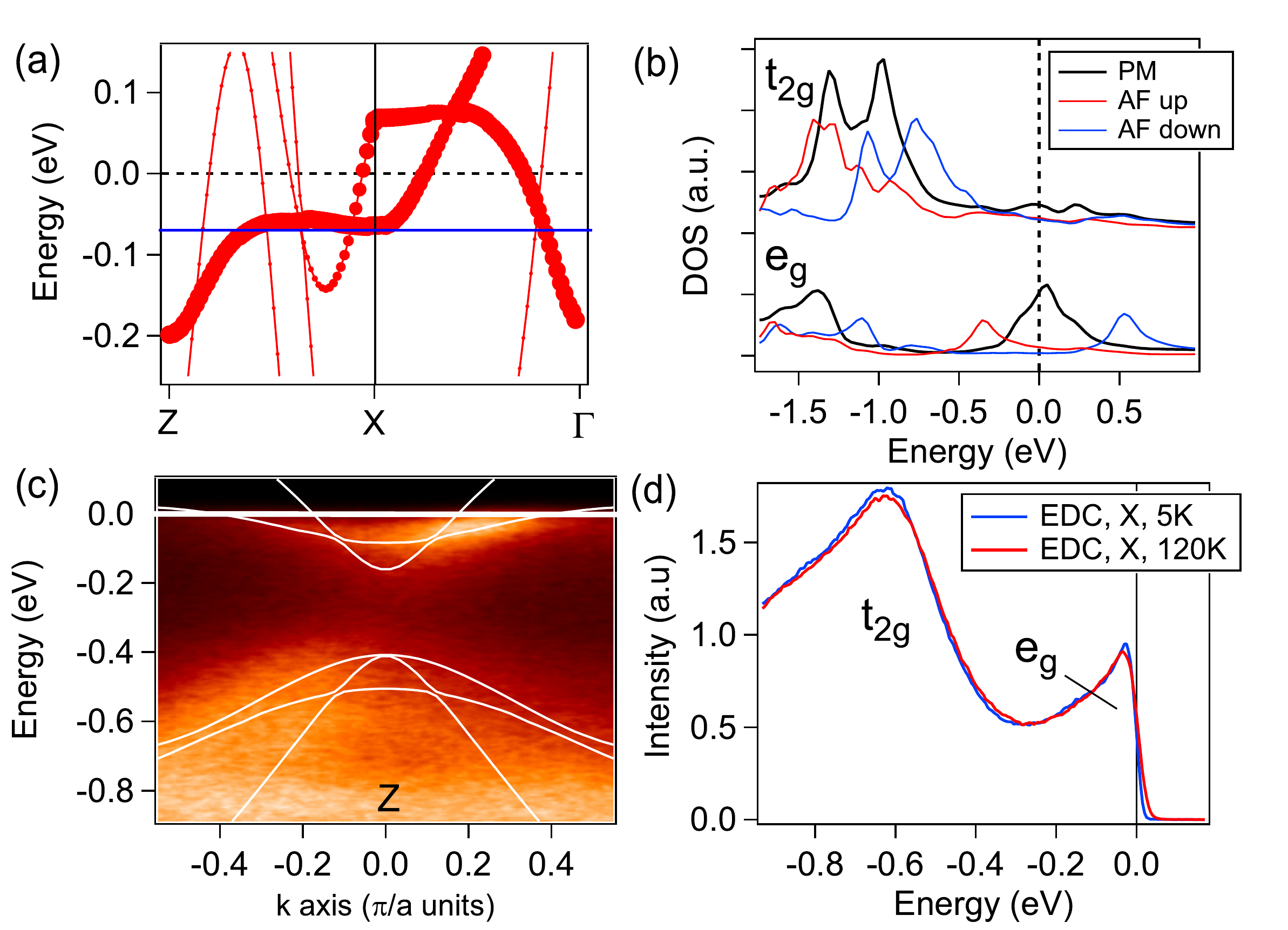}
\caption{(a) Calculated bulk dispersion for CaCo along ZX$\Gamma$. The $e_{g}$ weight is shown by the size of the markers. The calculation was neither shifted nor renormalized. The blue line indicates the position of $E_F$ found experimentally. (b) Density of states from the previous calculation in the paramagnetic (PM) phase and in the antiferromagnetic (AF) phase. (c) Energy-momentum image around Z of the small $d_{x^2-y^2}$ pocket. White lines are bulk calculation at $k_z$=1, renormalized by 1.4 and shifted up by 70~meV. (d) EDC spectra at X, showing the density of states for the $e_{g}$ pocket at 5K and 120K.}
\label{eg}
\end{figure}
We now investigate whether these differences can modify the electronic properties. As recalled in introduction, CaCo undergoes an AF transition at 72K in the bulk, where CoAs layers order ferromagnetically. In agreement with previous calculations \cite{AnandPRB14_Ca}, we find that there is a nearly flat band with $e_{g}$ character along $\Gamma$X/ZX Fig. \ref{eg}(a), which gives a high n($E_F$) in Fig. \ref{eg}(b). With the experimental magnetic structure, the calculation converges towards M=0.6$\mu_B$, consistent with the moments known experimentally \cite{QuirinalePRB13}. This splits the $e_{g}$ states near $E_F$ by nearly 0.5eV. However, in our ARPES data, we did not observe such a splitting. In Fig. \ref{eg}(d), we observe no difference for the $e_{g}$ peak at X at 5K or 120K.

Experimentally, we clearly see a small $e_{g}$ pocket at the zone center [Fig. \ref{eg}(c)], like in all Co pnictides. It is in good agreement with the calculation renormalized and shifted up by 70~meV (white lines). We further see another $e_{g}$ pocket at X, which is a fingerprint of the collapsed phase (compare FS of CaCo and SrCo in Fig. \ref{SrCo} and \ref{CaCo}). Indeed, in the maps of Fig. \ref{Periodicity}, the intensity at X is due to this pocket and clearly increases for the collapsed phases. This small pocket can be seen in the dispersion images of Fig. \ref{CaCo}(b) or (f). However, the band barely touches $E_F$ and it is difficult to extract its dispersion. The width of the EDC at X [Fig. \ref{eg}(d)] is about 50~meV. In the calculation, a larger pocket would be expected with a bottom at -0.1eV. However, if we apply the same renormalization of 1.4 and upward shift of 70~meV that we used for $d_{xz}$/$d_{yz}$ (the blue line in Fig. \ref{eg}(a-b) indicates the position of $E_F$ after such a shift), this band moves very close to $E_F$, in better agreement with the data. 

The 70~meV shift strongly reduces the $e_{g}$ weight and it is likely the reason for the suppression of the magnetic instability. Indeed in calculations for BaCo and SrCo, which do not order magnetically, the essential difference is the position of $e_{g}$ states, which are moved just above $E_F$. A careful examination of correlations in BaCo, including screened exchange beyond the local density approximation, also showed that the DFT calculation is not sufficient to correctly predict the position of this band \cite{VanRoekeghemPRL14}. 

\section{Discussion}
The electronic structure measured by ARPES in Co pnictides presents two \lq\lq{}anomalies\rq\rq{} compared to band structure calculations. (i) - a doublet is often measured, where a single line would be expected. The two lines of this doublet correspond quite well to the calculated extrema of the $k_z$ dispersion. (ii)- shifts as large as 0.2eV are required to match the bulk calculation in some locations of the BZ (Fig. \ref{SrCo}), while, in other parts, no shifts are required at all. 

We notice that both these features are present in a 2 layers slab calculation. Each line splits as a result of the interactions between the two layers, which \lq\lq{}replaces\rq\rq{} the $k_z$ dispersion of bulk systems. The locations in the reciprocal space where band shifts are observed in ARPES are also those where the deviation between the bulk and 2 layers slab calculations are the largest. This is a simple way to identify the places where the hybridization with bands having a strong 3D dispersion - the As $p_z$ band for example - is very large. 

When there are deviations between ARPES and bulk properties, it is generally assumed that there is a surface problem, for example a surface reconstruction. In ARPES studies of Fe pnictides, it is often argued that LiFeAs will be a \lq\lq{}better\rq\rq{} system for ARPES than AFe$_2$As$_2$, because its surface is non polar \cite{BorisenkoLiFeAs}. While this may be true, we emphasize that the problem we describe here is not due to a modification of the surface, but to an improper 3D dispersion. In fact, we have tried many different ways to modify theoretically the surface (through structural relaxation, charge unbalance...), but we could never reproduced the \lq\lq{}anomalies\rq\rq{} described above. As a consequence, systems like LiFeAs or FeSe, which are quite strongly 3D, could also be affected by such problems. 

From a practical point of view, this study gives indications on the situations where caution has to be taken to interpret ARPES data. 

(i) - One has to be aware that the 3D dispersion may modify the lineshapes in a way mimicking two different lines. For a complicated electronic structure as that of Fe pnictides, it could be tempting to attribute these lines to two different bands. 

(ii) - ARPES is often used to estimate the strength of electronic correlations from band renormalizations. As long as only renormalization of the band structure is used, this estimation should be reliable. However, shifts often have to be applied in Fe pnictides \cite{YiPRB09}, as a result of Fermi Surface \lq\lq{}shrinking\rq\rq{} \cite{OrtenziPRL09} or correlation-induced redistribution of carriers between orbitals \cite{LeePRL12}. In this case, the two quantities are not independent and defining a good renormalization may become a very tricky problem. For example, the bottom of the electron bands in these Co pnictides is around -0.85eV in the unrenormalized calculation and -0.6eV in the experiment. To match the two values, one could a priori renormalize by $\sim$1.4 or shift by $\sim$0.25eV, which have very different physical meaning. The changes in the $k_F$ contour as a function of $k_z$ should normally allow to choose the right solution, but, if there are intrinsic deviation between the bulk and calculated 3D dispersion, as we describe here, this may become impossible. One should keep in mind that a renormalization value will be reliable only if the $k_F$ vs $k_z$ contour has been determined in details.

(iii) - The most interesting consequence may be about the lineshape analysis. Linewidths should describe the lifetime of quasiparticles (QP) if no extrinsic source of broadening is present. In a 3D system, the 3D dispersion is often a major source of broadening that usually prevents lifetime analysis. It is clear from Fig. \ref{VsKz} that such analysis will be unfeasible here for a large $k_z$ range. However, we also observe rather clean lineshapes near $k_z$=1. In BaCo, they are even the sharpest lineshapes ($\Delta\nu$=0.03\AA$^{-1}$) that we have ever observed in our studies of these systems. It may be that the lifetime of the final state cancels at some points of the BZ, as observed in bulk Ag \cite{HansenPRL98}, making them favorable for such analysis. In a recent study of lifetimes in Ba(Fe,Co)$_2$As$_2$ and LiFeAs, we indeed concluded that the 3D brodening at favorable $k_z$ positions was not a major contribution to the linewidth. A more complete theory of the 3D broadening of these lineshapes would definitely be very interesting. 

\section {Conclusion}
In this study, we observed how increasing inter-layer interactions modify the electronic structure of Co pnictides. We show that ARPES largely follows this evolution, typical fingerprints of the \lq\lq{}collapsed\rq\rq{} electronic structure are for example observed in CaCo$_2$As$_2$. However, some deviations arise from the truncated 3D dispersion at the surface that are already significant for SrCo$_2$As$_2$. They are summarized in more details in the discussion section. We show that they should be carefully considered to give reliable estimations of the renomalization and/or the QP lifetimes from ARPES. We suggest that comparing calculations for the bulk structure and a 2 layers slab gives a good idea of places in the BZ where problems might appear. These deviations can have significant impact on the electronic properties. In CaCo$_2$As$_2$, they suppress the magnetic transition observed in the bulk. 

One interesting result of this study is that correlations do not change through the collapsed transition. In CaFe$_2$As$_2$, a change of correlations is expected at the collapsed transition \cite{DiehlPRB14,VanRoekeghemPRB16}, because of the disappearance of local magnetic moments \cite{YildirimPRL09,OrtenziPRL15,GretarssonPRL13}. This is partially supported by ARPES data \cite{XuFengPRB14,GofrykPRL14,DhakaPRB14}. The absence of change observed here is consistent with the idea that local moments have disappeared in Co pnictides, so that no change should be expected at the transition to the collapsed phase from this point of view. We note, nevertheless that the \lq\lq{}residual\rq\rq{} renormalization of 1.4 we find here is not insignificant. In fact, magnetism has not disappeared in Co pnictides, but the nature of magnetism has evolved, from a situation dominated by local moments to a much more itinerant picture. The correlation inherent to this state are also interesting. The value we find here is in good agreement with the enhancement of the susceptibillity and of Sommerfeld coeffecient quoted in these systems \cite{SefatPRB09,PandeyPRB13}. As n($E_F$) depends very sensitively on the position of the $e_{g}$ states compared to $E_F$, which may be slightly different between the bulk and the surface, more work will be needed to completely characterize the evolution of properties of this interesting family of materials. 

\vspace{0.5cm}

\textbf{ACKNOWLEDGMENTS}
We thank Silke Biermann for interesting discussions. This work was supported by the ANR “PNICTIDES”.

\bibliography{AllCo_Bib}

\end{document}